\documentclass[twocolumn]{aastex631}

\usepackage{color}

\usepackage{enumitem}
\usepackage{amsmath}
\usepackage{hyperref}

\newcommand{\be}{\begin{equation}}
\newcommand{\ee}{\end{equation}}

\newcommand{\lcdm}{\ensuremath{\Lambda\mathrm{CDM}}}

\providecommand{\sorthelp}[1]{}

\begin{document}

\title{Does the Correlation between 2MRS Galaxies and the CMB Indicate an Unmodeled CMB Foreground?}

\shorttitle{2MRS-CMB Correlation}
\shortauthors{G.~E.~Addison} 

\author[0000-0002-2147-2248]{Graeme~E.~Addison}
\correspondingauthor{Graeme~E.~Addison}
\email{gaddison@jhu.edu}
\affiliation{The William H. Miller III Department of Physics and Astronomy, Johns Hopkins University, 3400 N. Charles St., Baltimore, MD 21218-2686}

\begin{abstract}

\noindent
We revisit the claimed detection of a new cosmic microwave background (CMB) foreground based on the correlation between low-redshift 2MASS Redshift Survey (2MRS) galaxies and CMB temperature maps from the Planck and WMAP missions. We reproduce the reported measurements but argue that the original analysis significantly underestimated the uncertainties. We cross-correlate the 2MRS galaxy positions with simulated CMB maps and show that the correlation measured with the real data for late-type spiral galaxies at angular scales $\theta\geq0.1^{\circ}$ and redshift $cz<4500$~km~s$^{-1}$ is consistent with zero at the $1.7\sigma$ level or less, depending on the exact CMB map and simulation construction. This was the sample that formed the basis for the original detection claim.  For smaller angular separations the results are not robust to galaxy type or CMB cleaning method, and we are unable to draw firm conclusions. The original analysis did not propose a specific, falsifiable physical correlation mechanism, and it is impossible to rule out any contribution from an underlying physical effect. However, given our calculations, the lack of signal from expanding the redshift range, and the lack of corroboration from other galaxy surveys, we do not find the evidence for a new CMB foreground signal compelling.
\end{abstract}

\keywords{\href{http://astrothesaurus.org/uat/322}{Cosmic microwave background radiation (322)}; \href{http://astrothesaurus.org/uat/1146}{Observational cosmology (1146)}}

\section{Introduction}
\label{sec:intro}

Measurements of the CMB fluctuations, particularly from the Wilkinson Microwave Anisotropy Probe \citep[WMAP;][]{bennett/etal:2013} and Planck \citep{planck2016-l06} satellite missions, are a cornerstone of modern cosmology. The public data products from these experiments have seen extensive scrutiny within the community, motivated in-part by seemingly anomalous features in the data \citep[e.g.,][and references therein]{bennett/etal:2011,schwarz/etal:2016,planck2016-l07}, and ongoing discrepancies between the CMB predictions within the standard \lcdm\ model and other observations \citep[e.g.,][]{riess/etal:2022}. Obtaining cosmological constraints requires separating the CMB from astrophysical foreground microwave signals, and accurate foreground separation has become increasingly important as measurement precision has improved.

Recently, \citeauthor{luparello/etal:2023} (2023; hereafter L23) reported a detection of a new, previously unmodeled CMB foreground signal by correlating the positions of low-redshift galaxies ($cz\leq4500\textrm{~km~s}^{-1}$; $z\leq0.015$) in the 2MASS Redshift Survey \citep[2MRS;][]{huchra/etal:2012} with CMB maps provided by the Planck and WMAP teams. They showed that the thermal Sunyaev Zel'dovich and Integrated Sachs-Wolfe effects, known physical mechanisms that produce correlations between large-scale structure and the CMB, cannot explain this signal. If L23 have measured a genuine correlation and discovered a new foreground this could have wide-reaching implications. It may necessitate revising values of cosmological parameters estimated from the CMB, which are used very widely in modern cosmological and astronomical analyses. It could also have important implications for the interpretation of large-scale CMB anomalies, as discussed by \cite{hansen/etal:2023} and \cite{lambas/etal:2024}. 

In this Letter we revisit the L23 analysis in an attempt to better understand the origin of the reported correlation. We  reproduce their main results for elliptical and spiral galaxies in Section~\ref{sec:reproduce}, address the key issue of estimating uncertainty in the correlation statistic in Section~\ref{sec:statistics}, and provide concluding remarks in Section~\ref{sec:conclusions}.

\section{Computing the 2MRS-CMB Correlation}
\label{sec:reproduce}

Here we summarize the calculations performed by L23 and show that we can adequately reproduce them from public data. We follow L23 and use the SMICA temperature map and mask provided in the 2018 Planck data release\footnote{Specifically, we use the PR3-2018 Full Mission SMICA CMB map available from the Planck Legacy Archive at \url{https://pla.esac.esa.int/\#maps}.}. For the various 2MRS galaxy samples we use the public catalog\footnote{Specifically, we use the catalog file `2mrs\_1175\_done.fits' in the main catalog tar archive available from the 2MASS website \url{http://tdc-www.cfa.harvard.edu/2mrs}.} described by \cite{huchra/etal:2012}.

The angular correlation statistic used by L23 is (their Equation 1)
\be
\label{eqn:corrfunc}
\left\langle\Delta T(\theta)\right\rangle=\frac{1}{N_{\rm gal}}\sum_{k=1}^{N_{\rm gal}}\left(\frac{1}{N_{\textrm{pix},k}}\sum_{i\in C_k}\Delta T(k,i)\right),
\ee
where $i$ and $k$ label the CMB map pixels and galaxies, respectively, and $C_k$ is the set of $N_{\textrm{pix},k}$ pixels lying within the annulus with radius between $\theta$ and $\theta+\delta\theta$ from galaxy $k$. We exclude pixels that are masked in the SMICA confidence mask from the sum and adjust $N_{\textrm{pix},k}$ accordingly. Hereafter we drop the expectation brackets and denote the correlation statistic simply as $\Delta T(\theta)$. To approximately match the angular binning shown in the L23 figures we use logarithmically spaced angular bins with radii from $0.01^{\circ}$ to $20^{\circ}$.

\begin{figure}
\includegraphics[]{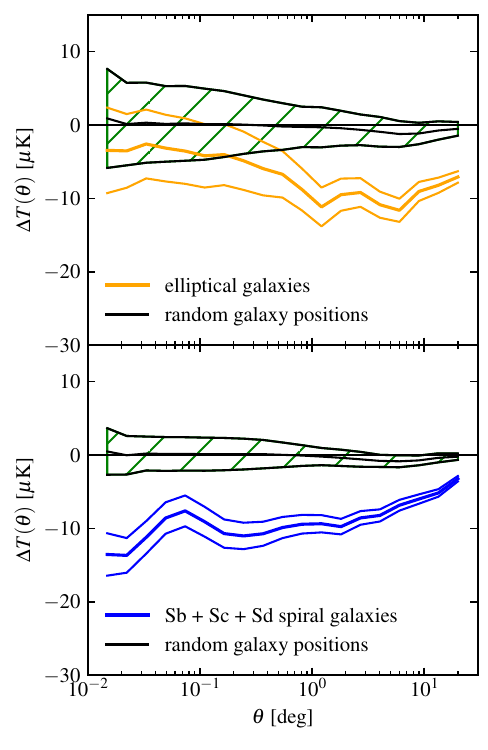}
\caption{Correlation between subsets of 2MRS galaxies at $cz<4500$~km~s$^{-1}$ and the Planck SMICA CMB temperature map. The top and bottom panels can be directly compared with panels (a) and (e) of Figure~2 of L23. We show two sets of uncertainty bands. The black lines and hashed regions correspond to the mean and standard deviation from mock galaxy samples with randomized positions. The colored bands correspond to the standard deviation across the real galaxy sample for each angular bin. The spirals in particular appear to exhibit a significant negative correlation over a wide range of scales, deviating from zero by many times the width of either set of uncertainty bands.}
\label{fig:reproduce}
\end{figure}

Figure~\ref{fig:reproduce} shows the correlation statistic we computed for the elliptical and late-type (Sb, Sc, and Sd class) spiral galaxies in the redshift range $300\textrm{~km~s}^{-1}\leq cz \leq4500\textrm{~km~s}^{-1}$. These results can be directly compared to panels (a) and (e) of Figure~2 of L23. In addition to the mean correlation, we compute the standard deviation of the correlation statistic across each galaxy sample and plot the mean plus and minus the standard deviation as the upper and lower colored lines in each panel, again following L23. We also computed the correlation using the real SMICA CMB map and mask but 100 simulated galaxy catalogs consisting of random, unclustered galaxy positions with the same number of entries as the real 2MRS data. The mean and standard deviation across the random galaxy samples are included as solid black lines in Figure~\ref{fig:reproduce} (cf. black lines and green hashed regions in Figure~2 of L23).

We reproduce the essential features of the L23 analysis: there is a seemingly highly significant detection of negative $\Delta T(\theta)$ on angular scales $1^{\circ}\lesssim\theta\lesssim20^{\circ}$ for both the elliptical and spiral galaxies for $cz<4500$~km~s$^{-1}$, given the uncertainties estimated from either (1) the standard deviation across the 2MRS galaxies for each annular bin, or (2) the spread in the mock galaxy samples with random positions. This negative $\Delta T$ extends to smaller scales for the spirals but not for the ellipticals. There are some differences between our Figure~\ref{fig:reproduce} and Figure~2 of L23, particularly for smaller angular separations ($\theta\lesssim0.1^{\circ}$). This may arise, at least in part, from differences in angular bin boundaries, or how pixels lying at bin boundaries are being assigned. As discussed in the next Section, results at $\theta<0.1^{\circ}$ also show more dependence on choice of CMB map. The more puzzling result from L23, and the primary motivation for our analysis, is the behavior at larger scales.

\section{Statistical significance of correlation}
\label{sec:statistics}

From a cosmological viewpoint, both the positions of the 2MRS galaxies, which trace the underlying density field, and the temperature fluctuations of the CMB, are random quantities that contribute scatter to the 2MRS-CMB correlation measured by L23. When comparing the data to simulations featuring no statistical correlation, one could generate synthetic versions of both the 2MRS catalogs and CMB skies. On average, however, the same correlation function uncertainty should be obtained from holding fixed one of the data sets, and only generating synthetic versions of the other\footnote{The L23 correlation $\Delta T(\theta)$ is directly related to the usual two-point angular correlation function. For uncorrelated fields $a$ and $b$ the variance of the two-point correlation depends on a four-point product of the form $\left\langle a^2\right\rangle\left\langle b^2\right\rangle$. One of $\left\langle a^2\right\rangle$ and $\left\langle b^2\right\rangle$ can be replaced with a fixed (measured) $a_0^2$ or $b_0^2$ without biasing the variance estimate.}. This was the approach taken by L23, who correlated synthetic galaxy catalogs against the (fixed) CMB data map. However, they used random, unclustered galaxy positions, whereas the distribution of the real 2MRS galaxies is highly anisotropic on the sky. This is clear by eye from Figure~4 of L23, and arises from a combination of the clumpy nature of the underlying density field and any non-uniformity in the selection  of the 2MRS sample.

Given the complexities of the real 2MRS data processing and selection, we do not have a straightforward way to generate more realistic, clustered, synthetic catalogs. We therefore considered a simpler method to estimate the $\Delta T$ uncertainties: correlating the real 2MRS data against simulated CMB temperature maps. We note that L23 did also use simulated CMB maps in their analysis (see their Section~4.2), but not for assessing detection significance.

One might wonder why it is informative to vary the CMB map given the consistency of results from switching out the Planck map for a WMAP-derived one. While the consistency of the Planck and WMAP results shown by L23 rules out a systematic origin specific to one of the CMB data sets (see their Appendix~B), here we are instead trying to quantify the scatter in $\Delta T$ expected from the cosmic variance in the CMB map (i.e., the different ways the harmonic mode amplitudes and phases could have been arranged while originating from the same underlying statistical distribution).

We generated Gaussian, statistically isotropic CMB temperature maps using the \texttt{synalm} routine in \texttt{healpy}\footnote{\url{https://healpy.readthedocs.io/en/latest/}}, the Python implementation of the \texttt{HEALPix} library \citep{gorski2005}. To produce maps with power consistent with the real SMICA maps we used the best-fit Planck 2018 \lcdm\ cosmological parameters from the joint fit to temperature, polarization and lensing power spectra \citep{planck2016-l06}. We then smoothed the simulated maps with a Gaussian beam with FWHM of $5'$, matching the SMICA processing \citep{planck2016-l04}. To improve agreement with the small-scale power in the real SMICA map (multipole moment $\ell>1500$) we finally added a white noise component with power $C_{\ell}=1.5\times10^{-4}$~$\mu$K$^2$.

\begin{figure}
\includegraphics[]{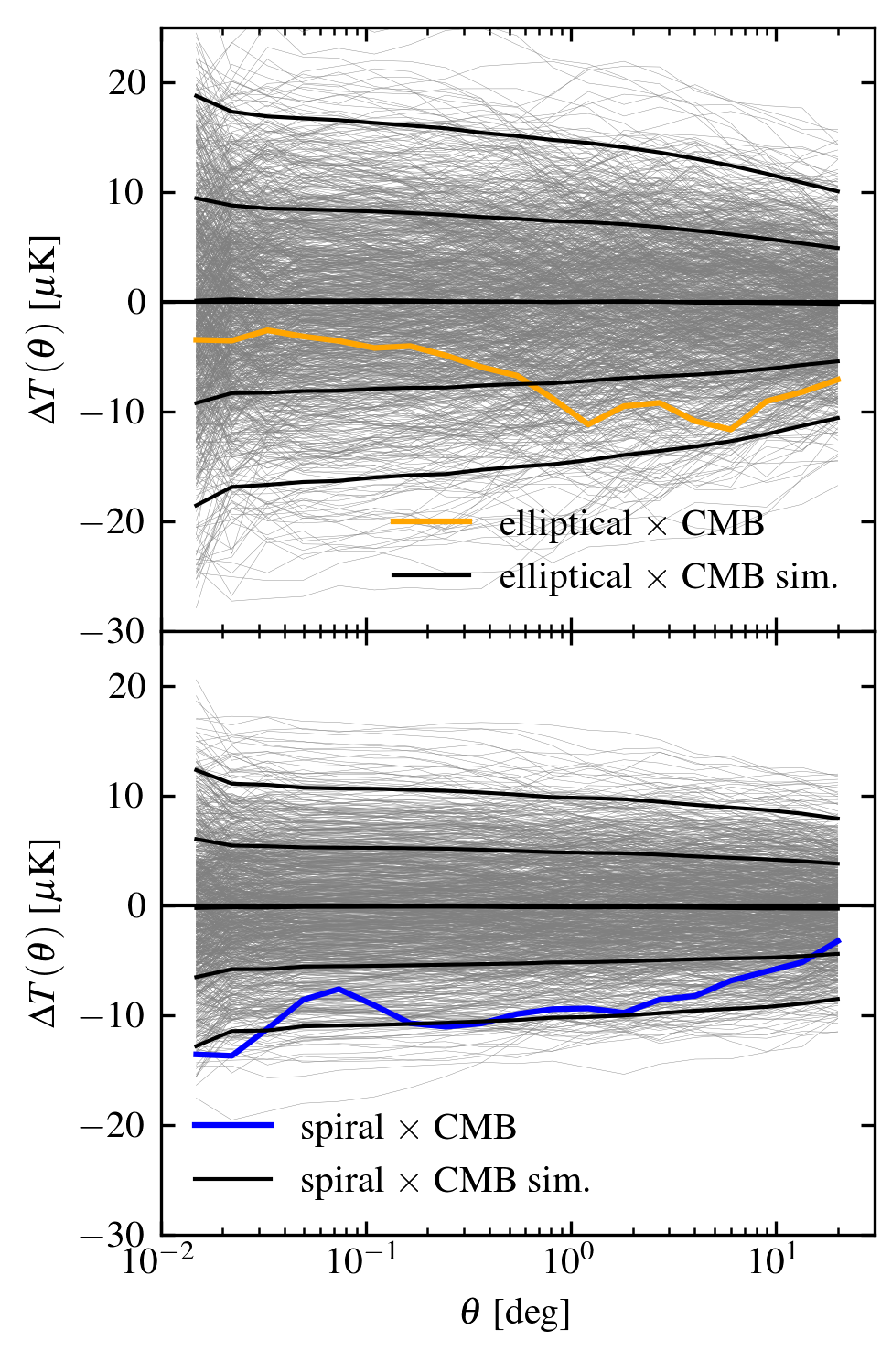}
\caption{Correlation between 2MRS galaxy samples at $cz<4500$~km~s$^{-1}$ and real or simulated CMB maps. The orange and blue lines are identical to those in Figure~\ref{fig:reproduce}. The grey lines show results using 1000 simulated CMB maps with no statistical correlation with the 2MRS galaxies, and the black lines denote the mean and $\pm1\sigma$, $\pm2\sigma$ intervals for the simulation ensembles. For separations $\theta>0.1^{\circ}$ the elliptical and spiral galaxy data correlations are consistent with zero within $0.2\sigma$ and $0.8\sigma$ (see text). For the concatenated elliptical-spiral data vector the difference is $1.1\sigma$. Different $\theta$ bins are highly correlated, making a by-eye $\chi^2$ estimate using the per-bin scatter inaccurate.}
\label{fig:cmbsim}
\end{figure}

Figure~\ref{fig:cmbsim} shows $\Delta T(\theta)$ obtained from correlating the elliptical and spiral 2MRS samples used in Figure~\ref{fig:reproduce} ($cz<4500$~km~s$^{-1}$) with 1000 simulated CMB maps. We applied the same SMICA confidence mask used in the analysis of the data map. We show the individual simulation results in Figure~\ref{fig:cmbsim} to emphasize that there are very significant off-diagonal correlations between different $\theta$-bins. More quantitatively, we find that neighboring bins are up to $97-99\%$ correlated, depending on the exact 2MRS sample, while, for example, the widely spaced bins centred around $\theta$ of $0.1^{\circ}$ and $20^{\circ}$ are correlated at the $70-90\%$ level.

We construct a simple $\chi^2$ statistic of the form $(d-m)^T\cdot C^{-1}\cdot(d-m)$ where $d$ is the (real or simulated) $\Delta T(\theta)$ vector, $m$ is the model prediction, fixed to zero to correspond to the null case of zero underlying 2MRS-CMB correlation, and $C^{-1}$ is the inverse covariance matrix. We find that simply inverting the sample covariance from the simulations is sufficiently accurate given the number of realizations and $\theta$-bins. We recover the $\chi^2$ distribution with the expected number of degrees of freedom for all the 2MRS-CMB simulation results, which also confirms that the distribution of the $\Delta T$ statistic is sufficiently Gaussian at all $\theta$.

On the real data, we find that the $\chi^2$ value for the 2MRS elliptical sample at $cz<4500$~km~s$^{-1}$ is consistent with the simulations, with a probability-to-exceed (PTE) of 0.64 based on the fraction of simulations with worse $\chi^2$. For the late-time (Sb, Sc, Sd) spirals, we find that the $\chi^2$ is far higher than in any of the simulations, but that this discrepancy is driven entirely by the correlation pattern at $\theta<0.1^{\circ}$. While the measured correlation clearly lies within the envelope of simulated results in Figure~\ref{fig:cmbsim}, the bin-to-bin variation is significantly different from that predicted in the simulations over this range. It is not clear what drives this behavior, although we have verified that the high $\chi^2$ is not from any single bin. Overall the results at $\theta<0.1^{\circ}$ are not fully consistent with galaxy type (see also Figure~2 of L23), or CMB map choice (Section~\ref{sec:sevem}, below). We therefore focus on results at $\theta>0.1^{\circ}$. In this case, the $\chi^2$ from the spiral sample is compatible with the simulations, with a PTE of 0.20 (corresponding to a $0.8\sigma$ deviation from $\Delta T=0$). We report PTE values for $\theta>0.1^{\circ}$ in Table~\ref{table:pte}.

The consistency between the elliptical and spiral galaxy correlations at $\theta\gtrsim1^{\circ}$ may seem to argue against a statistical fluke explaining the measured $\Delta T$. To test this, we formed a joint data vector, concatenating the $\Delta T(\theta)$ results from the two samples, and computed the $\chi^2$. These results are shown in Table~\ref{table:pte}. Combining the $\Delta T(\theta)$ measurements only slightly reduces the PTE for the SMICA CMB map compared to spirals alone (0.20 to 0.14). Based on the simulations, it is not in fact surprising to see similar correlation patterns with the elliptical and spiral samples. This is again due to strong off-diagonal covariance elements for $\Delta T(\theta)$, including in this case between the two galaxy samples, as clearly suggested by eye in their spatial distribution (Figure~4 of L23).

L23 performed further splits of the 2MRS sample by galaxy and environment properties. We have not investigated all of these, however we did perform calculations for the large spiral sample (with physical radius greater than 8.5~kpc, see Section~2.2 of L23). These results are shown in Table~\ref{table:pte}. This was the sub-sample with seemingly the strongest correlation signal, and L23 reported a $\geq4\sigma$ deviation from zero in each of the ten angular bins from $1^{\circ}$ to $10^{\circ}$ in their Section~5.  Overall, none of the calculations we have performed indicate deviations at even the $2\sigma$ level compared to results from correlating against the simulated CMB maps for $\theta>0.1^{\circ}$.

\begin{table}
\centering
\caption{Probability-To-Exceed (PTE) Simulated $\chi^2$ Values and Equivalent Gaussian ``$N\sigma$'' for 2MRS-CMB Correlations Measured at $\theta>0.1^{\circ}$ and Galaxy Redshift $cz<4500$~km~s$^{-1}$}
\begin{tabular}{llcc}
\hline
2MRS Galaxy Sample&CMB Map&PTE&Equiv.\\ 
&&&Gaussian\\
\hline
\hline
Spiral (Sb, Sc, Sd)&SMICA&0.20&$0.8\sigma$\\
&SEVEM&0.043&$1.7\sigma$\\
\hline
Elliptical&SMICA&0.43&$0.2\sigma$\\
&SEVEM&0.38&$0.3\sigma$\\
\hline
Spiral+Elliptical&SMICA&0.14&$1.1\sigma$\\
&SEVEM&0.052&$1.6\sigma$\\
\hline
Spiral, Large ($>8.5$~kpc)&SMICA&0.18&$0.9\sigma$\\
&SEVEM&0.10&$1.3\sigma$\\
\hline
\end{tabular}
\label{table:pte}
\end{table}

\begin{figure}
\includegraphics[]{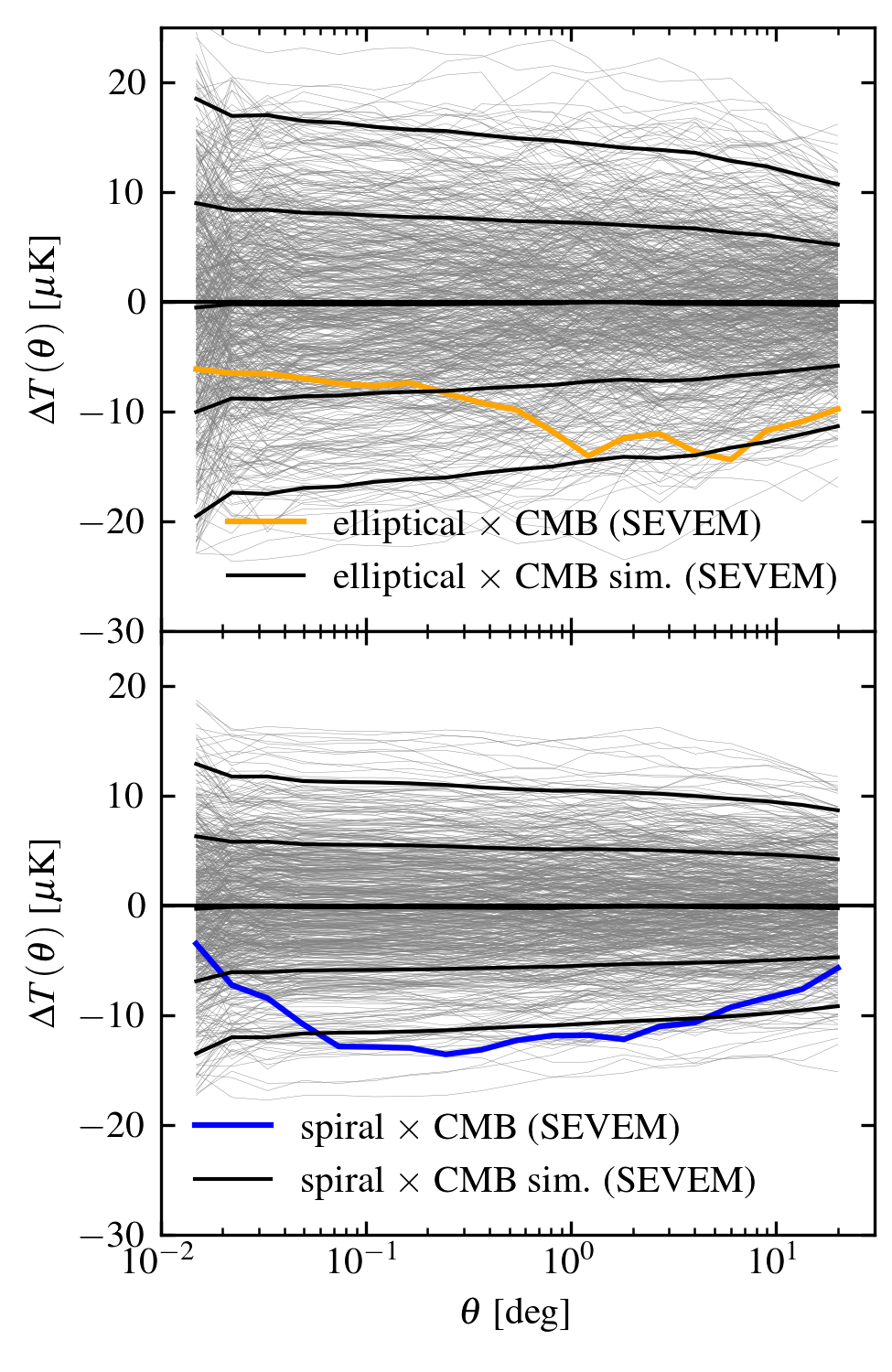}
\caption{Same as Figure~\ref{fig:cmbsim}, except using the Planck PR4/NPIPE SEVEM foreground separation method for the CMB maps, and the 600 simulated SEVEM-cleaned maps provided with the PR4 release. The SEVEM simulations include more realistic foreground and noise residuals than those used in Figure~\ref{fig:cmbsim}, however the $\chi^2$ for the spiral galaxies at $\theta<0.1^{\circ}$ remains a strong outlier from the simulations (see text). The data correlations at $\theta>0.1^{\circ}$, where results are more robust to CMB map choice and galaxy sample, are consistent with zero at the $0.3\sigma$ and $1.7\sigma$ levels for the ellipticals and spirals, or $1.6\sigma$ for the joint case.}
\label{fig:sevem}
\end{figure}

\subsection{Checking results with more realistic SEVEM CMB simulations}
\label{sec:sevem}

While our simplified CMB simulations were constructed to roughly match the power in the real SMICA CMB map, they do not contain realistic foreground residuals or noise. We investigated whether this could be driving the anomalous results for the spiral galaxies at small $\theta$. The Planck 2018 data release did not include a full set of SMICA-cleaned simulated maps (i.e., from applying the SMICA algorithm to simulations containing CMB, foregrounds and noise). However, 600 such simulations were provided for the alternative SEVEM foreground separation method \citep{planck2016-l04} in the subsequent PR4/NPIPE release\footnote{These simulations are available from the National Energy Research Scientific Computing Center (NERSC) with a user account, see \url{https://crd.lbl.gov/divisions/scidata/c3/c3-research/cosmic-microwave-background/cmb-data-at-nersc/}.} \citep{planck2020-LVII}. We therefore downloaded the PR4/NPIPE SEVEM cleaned data map plus the SEVEM simulations, and performed the same $\Delta T(\theta)$ calculation as before, retaining the 2018 SMICA confidence mask. These results are shown in Figure~\ref{fig:sevem}.

While the $\Delta T(\theta)$ results with the SEVEM CMB map are not identical to those from SMICA, particularly for the spiral galaxies at small $\theta$, the overall results are qualitatively consistent. The spiral galaxy $\chi^2$ remains anomalously high, and this is again due to bins at $\theta<0.1^{\circ}$. This indicates that our more simplistic CMB simulations described above were not the main factor causing the high $\chi^2$. Excluding $\theta<0.1^{\circ}$, the SEVEM $\chi^2$ is consistent with the simulations at the $0.3\sigma$ and $1.7\sigma$ levels for the elliptical and spiral samples (PTE values are reported in Table~\ref{table:pte}). Concatenating the data vectors from the two samples does not produce a significant detection, with the combined $\chi^2$ consistent with the simulations at the $1.6\sigma$ level.

\subsection{Understanding the discrepancy with L23 uncertainties}
\label{sec:discrepancy}

The uncertainty bands plotted in Figures~\ref{fig:cmbsim} and \ref{fig:sevem} are many times wider than those shown by L23. As mentioned earlier (see Figure~\ref{fig:reproduce}), it is notable that L23 found reasonable agreement between two different methods for estimating the uncertainties: (1) internal standard deviation across galaxies in the real 2MRS samples, and (2) scatter from correlating randomly generated, unclustered galaxy positions with the real CMB map. Our results imply both these methods must significantly underestimate the true uncertainties, and here we investigate this in more detail.

The standard deviation in $\Delta T(\theta)$ across the galaxies in the sample would be an appropriate measure of uncertainty if pixels belonging to annuli around the galaxy positions were independent for a given $\theta$ (i.e., if each galaxy contributed a unique set of pixels to the sum in Equation~\ref{eqn:corrfunc}). For larger $\theta$, however, this is increasingly not the case, since the annuli overlap. The highly anisotropic nature of the distribution of the 2MRS samples on the sky (Figure~4 of L23) exacerbates this issue. More quantitatively, for the 2MRS elliptical galaxies at $cz\leq4500$~km~s$^{-1}$, we find that around 75\% of pixels in the correlation sum are unique for $\theta\simeq1^{\circ}$, decreasing to only 5-10\% for $\theta\geq10^{\circ}$. For the spiral galaxies, the sample is several times larger, and around 60\% of pixels are unique for $\theta\simeq1^{\circ}$, decreasing to 1-3\% for $\theta\geq10^{\circ}$. This suggests, independent of our results using simulated CMB maps, that the L23 uncertainties are increasingly underestimated going to larger $\theta$.

As mentioned earlier, the 2MRS galaxies are not uniformly distributed across the sky but cluster, and this was not taken into account when L23 constructed simulated galaxy catalogs based on random sky positions. To roughly estimate the degree to which this might cause an underestimate in $\Delta T(\theta)$ uncertainties, we correlated the real SMICA CMB map with 2MRS samples with randomly rotated galaxy coordinates\footnote{More specifically, we applied rotations in three dimensions on the sphere using the \texttt{spatial.transform.Rotation.random} function of the \texttt{scipy} Python library.}. This has the advantage of preserving the internal 2MRS clustering pattern without requiring an actual model of the clustering (power spectrum, etc.), while also not relying on simulated CMB maps. Unfortunately this method is not perfect, and one particular issue we found is that systematically more galaxies are `lost' to the SMICA sky mask in the rotated coordinates than in the true coordinates. This is simply because in the true coordinates the Galactic plane cut in the SMICA mask roughly matches the portion of the sky with fewest 2MRS galaxies. Ignoring this caveat, the uncertainties in $\Delta T(\theta)$ for the rotated galaxies are larger than those from the random, unclustered samples by a factor of roughly two for smaller scales $\theta\lesssim0.1^{\circ}$, and up to a factor of five or more for $\theta\gtrsim10^{\circ}$ for the late-time spirals, based on 1000 random galaxy coordinate rotations. 

While we regard our calculations using the simulated CMB maps as more robust, the calculations just described provide further support to our argument that both methods used by L23 significantly underestimate the true $\Delta T(\theta)$ uncertainties.

\section{Conclusions}
\label{sec:conclusions}

We have reproduced the main results for the correlation statistic $\Delta T(\theta)$ from the L23 analysis using elliptical and spiral 2MRS galaxies at $cz<4500$~km~s$^{-1}$ and the Planck SMICA CMB temperature map. For separations $\theta\gtrsim0.1^{\circ}$ we find negative $\Delta T(\theta)$, roughly consistent for both galaxy types, which was the basis for the L23 claim of a new foreground detection.

We cross-correlated the 2MRS positions with simulated CMB maps that had zero underlying correlation with the galaxies. We found that for separation $\theta>0.1^{\circ}$ the measured elliptical and spiral correlations are actually consistent with zero at the $0.2\sigma$ and $0.8\sigma$ level (or $0.3\sigma$ and $1.7\sigma$ for the alternative PR4/NPIPE SEVEM CMB maps and simulations). While the correlation pattern for the elliptical and spiral samples is similar on these scales, the simulations show that this is not particularly surprising, and a joint analysis does not show a significant deviation from zero signal ($1.1\sigma$ and $1.6\sigma$ for SMICA and SEVEM). We highlighted the strong off-diagonal covariance elements that must be accounted for in a quantitative goodness-of-fit assessment. We argued that the uncertainties estimated by L23 are significantly underestimated (Section~\ref{sec:discrepancy}), and give the impression of a significant signal detection.

On smaller scales, the correlation measured in the spiral galaxies is discrepant with the simulated results. The $\Delta T(\theta)$ results at $\theta<0.1^{\circ}$ are not robust between different galaxy samples (as already shown by L23), or for the different CMB maps we considered. We therefore do not interpret the discrepancy with the simulations as indicating a genuine signal detection, although what exactly drives this difference is not clear.

L23 did not propose a specific, falsifiable physical mechanism for the 2MRS-CMB correlation, and we cannot rule out a physical mechanism making some contribution to the $\theta>0.1^{\circ}$ measurement. Based on our results in this work, however, we view a statistical fluctuation as an adequate explanation for the apparent correlation. While we have focused here on the redshift range highlighted as providing evidence for the new foreground signal by L23, $0.001<z<0.015$, it should also be pointed out that the correlation signal is diluted, and shifts towards zero, when a broader redshift range is considered. See Appendix~A and Figure~A.1 of L23 for results for $z<0.04$, or $cz<12000$~km~s$^{-1}$. Furthermore, to our knowledge, no corroborating signal has been reported from any other low-redshift galaxy surveys. The 2MASS data have been extremely valuable for many analyses in the past two decades, however more recent surveys such as the 6dF Galaxy Survey \citep[6dFGS;]{jones/etal:2004} and Sloan Digital Sky Survey \citep[SDSS; e.g.,][]{gunn/etal:2006} have the advantage of providing better characterized selection functions for cosmology-focused correlation studies.

Finally, we note here that the kinematic Sunyaev-Zel'dovich effect \citep[kSZ;][]{sunyaev/zeldovich:1980}, the Doppler shifting of the CMB photons from bulk motion of free electrons, may make some contribution to the measured correlation. This effect has the same frequency scaling as the primary CMB temperature fluctuations. Typically, the kSZ effect would average to zero in a cross-correlation between galaxies and the CMB, without some additional weighting, because roughly half the galaxies have peculiar velocity vectors pointing towards us, while the others point away \citep[e.g.,][]{hand/etal:2012}. However, for the small volume and very low redshift considered here, some coherent bulk motion may produce a net positive or negative signal. We have not performed any quantitative test of this.\\

Code to reproduce calculations in this work is available in a Zenodo repository \citep{addison:2024}.

I would like to thank Chuck Bennett, Mark Halpern, Gary Hinshaw, and Janet Weiland for helpful discussions relating to this work, and for providing comments on a draft of the manuscript. This work was supported in part by NASA ROSES grant 80NSSC24K0625. This work was based on observations obtained with Planck (http://www.esa.int/Planck), an ESA science mission with instruments and contributions directly funded by ESA Member States, NASA, and Canada. I acknowledge the use of the Legacy Archive for Microwave Background Data Analysis (LAMBDA), part of the High Energy Astrophysics Science Archive Center (HEASARC). HEASARC/LAMBDA is a service of the Astrophysics Science Division at the NASA Goddard Space Flight Center. This research has made use of NASA's Astrophysics Data System Bibliographic Services.

\software{numpy \citep{harris/etal:2020}, scipy \citep{virtanen/etal:2020}, matplotlib \citep{hunter:2007}, astropy \citep{astropy:2013,astropy2018,astropy2022}, HEALPix \citep{gorski2005}}

\end{document}